\documentclass{ws-procsmod}

\def\etal {{\it et al.}}
\def\mn{{\mu\nu}}
\def\la{\lambda}
\def\al{\alpha}
\def\be{\beta}
\def\ga{\gamma}
\def\de{\delta}
\def\ep{\epsilon}
\def\fr#1#2{{{#1} \over {#2}}}
\def\prt{\partial}
\def\ta{\tau}
\def\et{\eta}
\def\pp#1#2{\fr{\prt p_{#1}}{\prt p_{#2}}}

\newcommand{\refeq}[1]{(\ref{#1})}
\newcommand{\beq}{\begin{equation}}
\newcommand{\eeq}{\end{equation}}
\newcommand{\bea}{\begin{eqnarray}}
\newcommand{\eea}{\end{eqnarray}}

\begin{document}

\title{CLASSICAL LAGRANGE FUNCTIONS FOR THE SME
\footnote{Presented at the Fifth Meeting on CPT and Lorentz Symmetry,
Bloomington, Indiana, June 28-July 2, 2010
}}

\author{N.\ RUSSELL}

\address{Physics Department, Northern Michigan University\\
Marquette, MI 49855, USA\\
E-mail: nrussell@nmu.edu}

\begin{abstract}
A technique is presented for finding the classical Lagrange function
corresponding to a given dispersion relation.
This allows us to study the classical analogue
of the Standard-Model Extension.
Developments are discussed.
\end{abstract}

\bodymatter

\section{Introduction}
The Standard-Model Extension,
or SME,
is a framework for Lorentz
violation
that is set up in the context of
effective field theory\cite{kp}
in flat\cite{dcak}
and curved\cite{akgravity}
spacetime.
In the fermion sector,
the SME is established by
systematically adding Lorentz-breaking operators
of increasing mass dimension to the
Lagrange density.
For the minimal SME,
the dimension-three operators
have coefficients
$a_{\nu}$,
$b_{\nu}$,
$H_{\mn}$,
and the dimension-four operators
have coefficients
$c_{\mn}$,
$d_{\nu}$,
$e_{\nu}$,
$f_{\nu}$,
and
$g_{\la\mn}$.
Numerous experiments have placed bounds on these coefficients
and ones from the other sectors of the SME.\cite{datatables}

The dispersion relation
\beq
F(m,a,b,c,d,e,f,g,H;p)=0
\label{dr}
\eeq
for a mass $m$ fermion propagating
in these background fields
is known.\cite{2001causality}
To find it,
one seeks plane-wave solutions
to the modified Dirac equation.
In the resulting equation, a $4\times 4$ matrix acting on a spinor
is required to vanish.
The dispersion relation is the necessary condition that the matrix have zero determinant.
In the limits of only
$b_{\nu} \neq 0$, and only $c_{\mn}\neq 0$,
it takes the forms
\bea
0 &=&
(-p^2+ b^2+m^2)^2 - 4(b \cdot p)^2 + 4 b^2 p^2
\, , \label{bdr}\\
0 &=&
p\, (\de + 2 c + c^T c)p - m^2
\, . \label{cdr}
\eea
In our conventions,
light has unit speed
and the metric has diagonal entries
$(+1,-1,-1,-1)$.
For the minimal SME, the fermion-sector dispersion relation
is at most quartic
in the four-momentum $p_\nu$.
In particular,
note that Eq.\ \eqref{bdr} is quartic,
and that Eq.\ \eqref{cdr} is quadratic.
The full function $F$ involves numerous contractions
among the SME coefficients,
and factorization is not straightforward.\cite{2010Colladay}

Dispersion relations also exist for classical point particles.
For a conventional particle of mass $m$,
the classical Lagrange function is
\beq
L= - m \sqrt{u_\nu u^\nu}
\, .
\label{L:conventional}
\eeq
The canonical momentum
is
\beq
p_\nu
=
- \fr{\prt L}{\prt u^\nu}
=
\fr{m u_\nu}{\sqrt{u\cdot u}}
\, ,
\eeq
and the corresponding dispersion relation
is found by eliminating the four-velocity:
\beq
p^2
= m^2
\, .
\eeq

This proceedings contribution addresses the question of finding
the classical Lagrange function
corresponding to the
quantum-mechanics-derived SME dispersion relation
\refeq{dr}.
Effectively,
this means we are seeking
a method of constructing
the classical Lagrange function
from a given dispersion relation.

\section{Finding the Lagrange function}
We seek a
Lagrange function $L$
that yields the spacetime coordinates
$x^\nu$
for a particle of mass $m$
as a function of a curve parameter $\la$.
To meet the basic SME requirement of
conserved energy and conserved linear momentum,
$L$ cannot depend on time $x^0$ or position $(x^1,x^2,x^3)$.
So, it can take the form $L=L(u^\nu,\la)$,
where $u^\nu \equiv d x^\nu/d\la$.
The trajectory is found by extremization
of the action,
which is the integral of
$L(u^\nu,\la) d\la$ along a path.
To ensure that the result
is independent of the choice of curve parameter $\la$,
$L$ must also have no explicit $\la$-dependence
and must be homogeneous of degree one in $u^\nu$.
Using Euler's theorem for homogeneous functions,
this implies that
\beq
L(u) = \fr{\prt L}{\prt u^\nu} u^\nu
\equiv
-p_\nu u^\nu
\, .
\label{homog}
\eeq
This equation shows that the Lagrange function can be found
if the canonical momenta $p_\nu(u)$ are known.

To establish a match between a classical point particle
and the quantum-mechanical analogue of a particle,
we require that the classical velocity $dx^j/dx^0$
is the same as the group velocity of a plane-wave packet.
Noting that the classical velocity
can be related to the four velocity in the chosen
curve parametrization,
\beq
\fr{dx^j}{d x^0} = \fr{dx^j}{d\la} / \fr{dx^0}{d\la}
= \fr{u^j}{u^0}
\, ,
\eeq
we find the group-velocity condition to be
\beq
\fr{u^j}{u^0} = - \pp 0 j
\, ,
\label{gpvel}
\eeq
where $j$ is an index for the three cartesian spatial directions.
The partial derivative on the right-hand side of Eq.\ \eqref{gpvel}
can be evaluated by implicit differentiation of the given dispersion relation.

We can now state a method for finding
the Lagrange function for a given dispersion relation.
The dispersion relation \eqref{dr}
and the group-velocity conditions \eqref{gpvel}
form a set of four equations in the 8 variables
$u^\mu, p_\nu$;
we must solve this system for the 4 variables $p_\nu$,
in terms of the four variables $u^\mu$
and then substitute into Eq. \eqref{homog}
to get the Lagrange function.

With sufficiently small Lorentz breaking,
solutions for $L$ must exist because
they can only be a perturbation of the conventional ones.

An alternative
is to consider the set of 5 equations \eqref{dr}, \eqref{homog}, and \eqref{gpvel}
in the 9 variables $u^\mu$, $p_\nu$, and $L$;
use 4 of them to eliminate $p_\nu$,
and consider the resulting polynomial equation in $L$
with $u^\mu$-dependent coefficients.
Of the multiple solutions,
only real ones can be relevant,
and they correspond
very roughly to particles, antiparticles,
and differing spin-like states.

\section{The $b_\nu$ background}
As an example, we consider the
case of the SME $b_\nu$ background.
The procedure discussed above
leads to an octic polynomial in $L$,
expressed here in factorized form:
\bea
0&=&
\left(-b^2 (b\cdot u)^2 + b^2 L^2 -m^2(b\cdot u)^2\right)^2
\times
\left(b^2 u^2 - (b\cdot u)^2 + (L + m \sqrt {u^2})^2 \right)
\nonumber\\
&&
\times \left(b^2 u^2 - (b\cdot u)^2 + (L - m \sqrt {u^2})^2 \right)
\, .
\eea
For vanishing $b_\nu$,
the second and third factors give
the expected limits of $L$,
which are Eq.\ \eqref{L:conventional}
and its negative-mass counterpart.
We disregard the first term as unphysical.
The solutions for the positive-mass case
are:
\beq
L = - m \sqrt{u^2} \mp \sqrt{(b\cdot u)^2 - b^2 u^2}
\, .
\label{L:b}
\eeq
The corresponding canonical momenta are
\beq
p_\nu =  \fr{m u_\nu}{\sqrt{u^2}}
\pm \fr{(b\cdot u) b_\nu - b^2 u_\nu}{\sqrt{(b\cdot u)^2 - b^2 u^2}}
\, .
\label{p:b}
\eeq
This reveals that the conserved 3-momentum $p^j$
is not collinear with the 3-velocity $u^j$,
and that the one can be nonzero when the other vanishes.
Another property
is the dependence of the conserved energy $E\equiv p_0$
on the direction of the 3-velocity.
This differs from the conventional case,
where the dependence is only on the magnitude of the 3-velocity.

Equation \eqref{p:b} is given in parametrization-independent form.
While the proper-time interval
along the curve
$d\ta=(\et_\mn dx^\mu dx^\nu)^{1/2}$
can always be chosen as the curve parameter,
other choices may be more convenient for different SME terms.
In this case, the proper-time choice
$d\la=d\ta$
leads to $u_\nu u^\nu = 1$,
and provides a convenient simplification.

\section{Spacetime torsion}
The spacetime torsion
tensor,\cite{HehlHammond}
denoted ${T^\mu}_{\al\be}$,
has a close relationship
to various terms in the SME.
To appreciate the rich structure
underlying its 24 independent components,
we may draw upon the standard $4+4+16$ irreducible decomposition
with components $A^\nu$, $T_\nu$, $M_{\la\mn}$.
The axial part $A_\nu$
can be defined by
\beq
A^\mu \equiv \fr 1 6 \ep^{\al\be\ga\mu}T_{\al\be\ga}
\, .
\eeq
If torsion is present,
this axial component
enters the Dirac equation
as a coupling to spin.
It allows limits to be placed based on
experiments seeking
spacetime anisotropies\cite{1997Laemmerzahl}
and Lorentz violation.\cite{2002Shapiro}

Recent work has investigated
nonminimally-coupled torsion
and its relation to the SME background fields.
Using results from experiments testing Lorentz symmetry,
the first limits were placed on
19 of the 24 independent
components of the torsion tensor.\cite{2008krt}
These are measured in
the standard inertial reference frame used for SME
experiments.\cite{datatables}
The experiments involved were
a dual-maser
system\cite{2004Cane,2008krt}
at the Harvard-Smithsonian Center for Astrophysics,
and a spin-polarized torsion pendulum
\cite{2008Heckel}
at the University of Washington in Seattle.

Using the conventions of Ref.\ \refcite{2008krt},
the correspondence between axial torsion and the $b_\nu$
background is
\beq
b_\nu = - \fr 3 4 A_\nu
\, .
\eeq
Exploiting this relationship
derived in the quantum-mechanical context,
we may deduce
from
Eq.\ \eqref{L:b}
the Lagrange function for a classical particle
in a constant minimal-torsion background.
For a point particle
of mass $m$
in Minkowski spacetime,
the result is
\beq
L = - m \sqrt{u^2} \mp \fr 3 4\sqrt{(A\cdot u)^2 - A^2 u^2}
\, .
\eeq
We remark that this Lagrange function is valid
at all orders in the axial torsion tensor.
We may readily deduce the dispersion relation and canonical momentum,
and the results are identical to those given in Eqs.\ \eqref{bdr}
and \eqref{p:b}, up to factors of $-3/4$.

\section{Discussion}
Our main result is a method that,
in principle,
can generate the classical Lagrange function
corresponding to
the full minimal-SME dispersion relation.
Finding the full Lagrangian
is technically difficult
because it involves solving a polynomial of high order.
A useful illustrative example is
the case of the $b_\nu$
background.
Using this,
we have deduced the Lagrange function for
a point particle in the presence of minimal torsion.
There are many other avenues for investigation,
some of which are discussed elsewhere.\cite{2003Lehnert,2005badc,aknr}

\end{document}